\documentclass[12pt,english]{article}
\ifx\pdfoutput\undefined
\usepackage[dvips]{graphicx}
\else
\fi
\usepackage[utf8]{inputenc}
\usepackage{blindtext}
\usepackage{epstopdf}
\usepackage{floatrow}
\usepackage{graphicx}
\usepackage{fancyhdr}
\usepackage{multirow} 
\usepackage{amsfonts}
\usepackage{amsmath}
\usepackage{amssymb}
\usepackage{natbib}
\usepackage{epsfig}
\usepackage{babel}
\usepackage{float}
\usepackage{soul}
\usepackage{bm}
\newfloatcommand{capbtabbox}{table}[][\FBwidth]

\floatstyle{plaintop}
\restylefloat{table}
\usepackage[font=small,labelfont=bf]{caption}
\usepackage[pdftex,dvipsnames,usenames]{color} 
     
\usepackage{sectsty}
\allsectionsfont{\usefont{OT1}{phv}{bc}{n}\selectfont}

\newcommand{\var}{\text{Var}}
\newcommand{\E}{\text{E}}

\begin{document}
\thispagestyle{empty}

\centerline{\bf{Comparison of Bayesian Nonparametric Density Estimation Methods}}
\vskip 3mm
\vskip 5mm
\noindent Adel Bedoui\textsuperscript{a}\footnote{Corresponding author: Adel Bedoui. Email: bedoui.adel1@gmail.com}
 and Ori Rosen\textsuperscript{b}
\vskip 5mm

\noindent \textsuperscript{a}Department of Statistics, University of Georgia, Athens, Georgia \ 30602 \\
\textsuperscript{b}Department of Mathematical Sciences, University of Texas at El Paso,  El Paso, Texas \ 79968

\vskip 3mm

\noindent ABSTRACT

In this paper, we propose a nonparametric Bayesian approach for Lindsey and penalized Gaussian mixtures methods. We compare these methods with the Dirichlet process mixture model. Our approach is a Bayesian nonparametric method not based solely on a parametric family of probability distributions. Thus, the fitted models are more robust to model misspecification. Also, with the Bayesian approach, we have the entire posterior distribution of our parameter of interest; it can be summarized through credible intervals, mean, median, standard deviation, quantiles, etc. The Lindsey, penalized Gaussian mixtures, and Dirichlet process mixture methods are reviewed. The estimations are performed via Markov chain Monte Carlo (MCMC) methods. The penalized Gaussian mixtures method is implemented via Hamiltonian Monte Carlo (HMC). We show that under certain regularity conditions, and as n increases, the posterior distribution of the weights converges to a Normal
distribution. Simulation results and data analysis are reported.

\vskip 4mm

\section{Introduction}
A hallmark of many modern data problems is to find the characteristics and the distribution of data. For instance, the objectives could be the extraction of information about skewness and multimodality in the data that are not available at first glance. As a result density estimation has become increasingly popular. In finance, density estimation plays an important role in detecting heavy tails and patterns in a market stock. A common classical approach is the histogram \citep{Pearson1895}. Several approaches to nonparametric density estimation are available, the most common of which is kernel estimation \citep{rosenblatt:1956,Parzen1962,nadaraya:1964,watson:1964,wand:jones:1995}. 
In this paper, we consider three less traditional approaches to nonparametric density estimation: Lindsey's method 
\citep{Lindsey1974}, Penalized Gaussian Mixtures (PGM) \citep{Ghidey2004} and Dirichlet Process Mixture Models (DPMM) \citep{Ferguson1973}. \cite{schellhase:kauermann:2012} estimate densities using penalized mixtures of $B$-spline densities
and compare it via simulation to other estimation methods, including kernel estimation and classical finite mixtures. \cite{schellhase:kauermann:2012} discuss
penalized mixtures in detail, and refer to {\tt R} packages that implement the methods to which penalized mixtures are compared. Unlike their approach, ours is Bayesian. Moreover, we implement the HMC to estimate weights in the PGM method and show that, under certain regularity, the posterior distribution of the coefficients converges to a Normal distribution. In addition, we derive a Bayesian version for Lindsey's method, which converts density estimation to a regression problem. To estimate the regression function, we applied the cubic smoothing spline as it is a flexible approach to fitting data such that it produces a smooth function. The use of the Bayesian nonparametric method has various advantages. For example, the resulting method is more robust to the data distribution, as it is partially based on the parameterized families of prior probabilities distributions. Moreover, we have the entire posterior distribution of our parameter of interest, which can be summarized through credible intervals, mean, median, standard deviation, quantiles, range, etc. That is,  probability comes into play in a Bayesian credible interval after collecting the data. For example, based on the data, we think there is a 95\% probability that the true parameter value is in the interval. However, the probability comes into play in a frequentist confidence interval before collecting the data. For instance,  we think there is a 95\% probability that we will collect data that produces an interval that contains the true parameter value.

The paper is organized as follows. Section~\ref{sec:poisson} introduces Lindsey's method.
Sections \ref{sec:mix_gauss} and \ref{sec:dpmm} describe the PGM and the DPMM approaches, respectively. A simulation study and an example with real data are provided in Section \ref{sec:example}. Section \ref{sec:summary} contains a summary.

\section{Lindsey's Method}
\label{sec:poisson}
Lindsey's method (LM) \citep{Lindsey1974} recasts density estimation as a regression problem.
\cite{Wasserman} describes it as follows. Suppose a random sample $X_1,X_2, \ldots, X_n$ from a density $f$ on [a,b] 
is available. Divide the interval [a,b] into $k$ equal width bins
and denote by $N_j$ and $t_j$ the number of observations and the abscissa of the center of the $j$th bin, respectively.
Let $Y_j=\sqrt{k/n}\times\sqrt{N_j+1/4}$, then 
\begin{equation}
Y_j\approx r(t_j)+\sigma\epsilon_j,
\label{eq:lindsey_reg}
\end{equation}
where $\epsilon_j\sim N(0,1)$, $\sigma=\sqrt{k/(4n)}$ and $r(x)=\sqrt{f(x)}$.
The regression model (\ref{eq:lindsey_reg}) follows from the assumption that
$N_j$ has an approximate Poisson distribution with mean $n\,f(t_j)/k$ and from applying the delta method.
More specifically, using $\E(N_j)=\var(N_j)\approx n\,f(t_j)/k$ and applying the delta method show
that $\E(Y_j)\approx\sqrt{f(t_j)}$ and $\var(Y_j)\approx k/(4n)$.
To estimate $f$, one applies their favorite nonparametric regression procedure to the data $(t_j,Y_j)$, $j=1,\ldots,n$,
to obtain $\hat{r}$. An estimate $\hat{f}$ is then given by
\[
\hat{f}(t)=(r^+(t))^2/\int_0^1(r^+(t))^2dt,
\]
where $r^+(t)=\max\{\hat{r}(t),0\}$. In this paper, we use the cubic smoothing splines to obtain $\hat{r}$.
Lindsey's method is described in many other references, for example \cite{brown:cai:zhang:zhao:zhou:2010}
and \cite{Efron2010}.
\subsection{Applying Cubic Smoothing Splines to Estimate the Regression Function}
We now estimate the function $r(t)$ in Equation (\ref{eq:lindsey_reg}) by the cubic smoothing splines method using \citet{Wahba1990}'s approach. Precisely, we implement a Bayesian framework to estimate $r(t)$. We let:
\[r(t)=\beta_0 + \beta_1t +h(t),\]
where $h=(h(t_1),h(t_2),.....,h(t_n))'$ is a zero-mean Gaussian process with variance covariance matrix $\tau^2 \Omega$, with the $ij^{th}$ element of $\Omega$  given by	
\begin{equation} 
\Omega_{ij}=\dfrac{1}{2} t^2_i(t_j - \dfrac{t_i}{3}),\; t_i<t_j.
\nonumber
\end{equation}
To facilitate the computation, we write $h=Z\bm{u}$ where $Z$ is obtained as follows. The matrix $\Omega$ is expressed as $\Omega =QDQ'$, where $Q$ is the matrix of eigenvectors of $\Omega$ and $D$ is a diagonal matrix containing the eigenvalues of $\Omega$. Letting $Z=QD^{1/2}$ and setting the prior on $\bm{u}$ to be N($0,\tau^2 I_n$) means that $Z\bm{u}\sim N(0,\tau^2\Omega)$. The parameter $\tau^2$ is a smoothing parameter controlling the smoothness of $r(t)$.
\subsubsection{Prior Distributions}
The following priors are placed on the parameters:
\begin{enumerate}
	\item $\bm{\beta} \sim N(\bm{0},\sigma_{\beta}^2I_2)$, $\sigma_\beta ^2$ is a large fixed number.
	\item $\bm{u} \sim N(\bm{0},\tau^2I_m).$
	\item $\tau^2 \sim U(0, c_{\tau^2})$, $c_{\tau^2} \text{ is fixed}$.
	\item $\sigma^2 \sim U(0, c_{\sigma^2})$, $c_{\sigma^2} \text{ is fixed}$.
\end{enumerate}
The model can thus be written as:
\begin{equation}
\bm{Y}= X\bm{\beta} +Z\bm{u}+\bm{\epsilon},
\nonumber
\end{equation} 
where
\begin{center} X=
	$\begin{pmatrix}
	1 & t_1\\ 
	1& t_2\\ 
	.& .\\ 
	.& .\\ 
	1& t_n
	\end{pmatrix}$.
\end{center}
To properly model $\bm{u}$, One should be careful in selecting the choice of $c_{\tau^2}$ and $c_{\sigma^2}$ in the uniform priors used for $\tau^2$ and $\sigma^2$. The eventual choice plays a critical role in the amount of smoothing of the data. One good choice is to let $c_{\tau^2}=10^5$ and $c_{\sigma^2}=10^3$.
The number of columns of $Z$ is reduced from $n$ to $m$ $(m < n)$ by retaining only the $m$ columns, corresponding to the $m$ largest eigenvalues of $\Omega$ without affecting the fit.
\subsubsection{Gibbs Sampling}
To sample from the posterior  $p(\bm{\beta},\bm{u},\tau^2,\sigma^2|\bm{Y},Z,X),$ we draw from the following conditional distributions
\begin{enumerate}
	\item Sample $\bm{\beta},\bm{u}$ from
	\[\begin{split}
	N\Bigl{(}  \frac{1}{\sigma^2}(\sigma^2(X^{*'}X^* + \sigma^2A^{-1})^{-1})X^{*'}\bm{Y} &,\; \sigma^2(X^{*'}X^* + \sigma^2A^{-1})^{-1}\Bigr{)}, 
	\nonumber
	\end{split}\]  
	where $X^*= \Bigl{(} X | Z \Bigr{)}$, i.e., where $X$ and $Z$ are concatenated columnwise, and $A=\mbox{diag}(\sigma_{\bm{\beta}}^2,\sigma_{\bm{\beta}}^2,\tau^2 ,\cdots,\tau^2)$
	\label{betau}
	
	\item Sample $\sigma^2$ from
	\begin{equation} 
	IG\Bigl{(} \frac{n}{2} -1 ,\frac{1}{2}(\bm{Y} - X^{*}\bm{\beta}^{*})^{'}(\bm{Y} -X^{*}\bm{\beta}^{*}\Bigr{)}\cdot I(0 \leq \sigma^2 \leq c_{\sigma^2}),
	\label{sigma}
	\end{equation}
	i.e., a truncated $IG$ distribution. $\bm{\beta}^*= \Bigl{(} \bm{\beta} | \bm{u} \Bigr{)}$, i.e., where $\bm{\beta}$ and $\bm{u}$ are concatenated columnwise
	\item Sample $\tau^2$
	\begin{equation}
	IG\Bigl{(}\frac{m}{2} -1, \frac{1}{2}\bm{u}^{'}\bm{u}\Bigr{)} . I(0 \leq \tau^2 \leq c_{\tau^2}). 
	\label{tau}
	\end{equation}
\end{enumerate}

\section{Penalized Gaussian Mixtures}
\label{sec:mix_gauss}
\cite{Ghidey2004} proposed a density estimation method based on a mixture of Gaussians with fixed parameters.
In particular, the interval $[a,b]$ is first divided into a grid of equally spaced points, $\mu_1,\mu_2,\ldots,\mu_K$,
$\mu_j<\mu_{j+1}$, which are used as the means of the Gaussian densities in the following mixture model
\begin{equation}
f(x)=\sum\limits_{j=1}^{K}c_j \dfrac{1}{\sqrt{2\pi \sigma^2}} \mbox{exp}\Bigl{\{} - \dfrac{(x-\mu_j)^2}{2\sigma^2}\Bigr{\}},
\label{eq:gauss_mix}
\end{equation}
where $\sigma=\frac{2}{3}(\mu_{j+1} - \mu_j)$. This choice of $\sigma$ is based on 
equating the width of the support of a cubic $B$-spline consisting of four equal intervals of width $(\mu_{j+1}-\mu_j)$,
each, to $6\sigma$. As an alternative to Gaussian densities, $B$-spline densities can be used as the basis
functions, see for example \cite{staudenmayer:ruppert:buonaccorsi:2008}.
As for the value of $K$, \cite{schellhase:kauermann:2012} recommend that it be large enough but usually small compared to the sample size, which is the rule of thumb advocated by \cite{ruppert:2002}.
The weights $c_j$ in (\ref{eq:gauss_mix}) are given by
\[
c_j= \frac{\exp (\beta_j)}{\sum_{h=1}^K \exp (\beta_h)},
\] 
such that $\sum_{j=1}^{K} c_j=1$, and $\beta_1$ is set to zero  for identifiability. 

Let $\bm{\beta}=(\beta_2,\ldots,\beta_K)'$. To obtain a smooth fit, the $c_j$ corresponding to 
neighboring Gaussian densities must be close to one another \citep{Eilers1996}. 
This can be achieved by constraining the corresponding $\beta_j$ to be close to one another.
By analogy to \cite{Eilers1996}'s idea, \cite{Lang2004} require the
$\beta_j$ to satisfy $\beta_\rho=2\beta_{\rho-1}-\beta_{\rho-2}+u_\rho$,
where $u_\rho\sim N(0,\tau^2)$, and $p(\beta_2)\propto 1$,
$p(\beta_3)\propto 1$. However, this results in an improper prior distribution
on $\bm{\beta}$ similar to the intrinsic Gaussian Markov random field prior used in 
spatial statistics. \cite{Chib2006} place a joint normal prior on
$\beta_2$ and $\beta_3$ which fixes the impropriety of the prior
on $\bm{\beta}$.
For example, if $(\beta_2,\beta_3)'\sim N_2(\bm{0},c\,\tau^2 I_2)$, where $c$ is a fixed
constant, the prior on $\bm{\beta}$ becomes
\begin{equation}
p(\bm{\beta} \mid \tau^2)\propto (\tau^2)^{-\frac{1}{2}(K-1)}
\exp\Bigl\{-\frac{1}{2\tau^2}\Bigl [\sum_{\rho=4}^{K}(\beta_\rho-2\beta_{\rho-1}+\beta_{\rho-2})^2 
+c^{-1}\bm{\beta}'_{2:3}\bm{\beta}_{2:3}\Bigr ]\Bigr\},
\label{eq:beta_prior}
\end{equation}
where $\bm{\beta}_{2:3}$ is a vector consisting of the first two entries of $\bm{\beta}$.
The summation in the exponent on the right-hand side of (\ref{eq:beta_prior})
can be expressed as $\bm{\beta}' P\bm{\beta}$,
where $P=D' D$ and $D$ is the $(K-3)\times (K-1)$ matrix
\[
\left (
\begin{array}{rrrrrrr}
1 & -2 & 1  & 0 & 0 & \ldots &     0   \\
0 &  1 & -2 & 1 & 0  & \ldots &    0    \\
\vdots   & \ldots    &     &    &     &    &  \vdots   \\
0 & 0 & \ldots  &  & 1 & -2 & 1 \\
\end{array}
\right).
\]
Equation~(\ref{eq:beta_prior}) can now be re-expressed as
\[
p(\bm{\beta} \mid \tau^2)\propto (\tau^2)^{-\frac{1}{2}(K-1)}
\exp\Bigl\{-\frac{1}{2\tau^2} \bm{\beta}' P^* \bm{\beta}\Bigr\},
\]
where $P_{ll}^*=P_{ll}+c^{-1}$, for $l=1,2$.

The parameter $\tau^2$ determines how smooth the estimated density will be.
We place a Half-$t(\nu,A)$ \citep{gelman:2006} distribution on $\tau$, whose pdf is
$p(x)\propto [1+(x/A)^2/\nu)]^{-(\nu+1)/2}$, $x>0$, where the hyperparameters $\nu$ and $A$ are
assumed known. The larger the value of $A$, the less informative the prior is. In addition, the weights, $c_j$ , are not sensitive to the value of A (see Appendix A3). Computationally, it is convenient to utilize the following scale mixture representation
\citep{wand:etal:2012}: $(\tau^2\mid a) \sim IG(\nu/2,\nu/a)$, $a\sim IG(1/2,1/A^2)$,
where $IG(a,b)$, is the inverse Gamma distribution with pdf
$p(x)\propto x^{-(a+1)}\exp(-b/x)$, $x>0$.

\subsection{Estimation}
Given a sample $X_1, X_2, ...,X_n$, it is convenient to augment the data with indicators $z_i$
taking values in $\{1,2,...,K\}$. The augmented likelihood is then given by
\begin{equation}
f(\bm{\beta}\mid \bm{x},\bm{z})= \prod_{i=1}^{n} c_{z_i}\frac{1}{\sqrt{2\pi \sigma^2}} \exp\Bigl\{- \frac{(x_i-\mu_{z_i})^2}{2\sigma^2}\Bigr\}.
\label{eq:aug_lik}
\end{equation}
Equation (\ref{eq:aug_lik}) in combination with the priors give rise to the following sampling scheme.
\begin{enumerate}
	\item
	Sample $\bm\beta$ from
	\begin{equation}
	p(\bm\beta \mid \tau^2,\bm z)\propto \prod_{j=1}^{K} c_j^{n_j}\exp \Bigl\{ - \frac{1}{2\tau^2}\bm{\beta}'P^{*}\bm\beta \Bigr\},
	\label{eq:cond_beta}
	\end{equation}
	where $n_j=\#\Bigl\{ z_i=j \Bigr\}$. A Hamiltonian Monte Carlo (HMC) algorithm is used to sample from this distribution. More details
	are given in the Appendix.
	\item
	Sample $\tau^2$ from $IG\Bigl((K+\nu-1)/2,\, \nu/a+\frac{1}{2}\bm\beta'P^*\bm\beta\Bigr)$.
	\item
	Sample $a$ from $IG\Bigl((\nu+1)/2,\, 1/A^2+\nu/\tau^2 \Bigr)$.
	\item
	Sample the indicators one at a time from multinomial distributions $M(1,h_{i1},...,h_{iK})$, where 
	\(
	h_{ij}= c_j\, \exp \bigl\{-\frac{1}{2}(x_i -\mu_j)^2/\sigma^2   \bigr\}/ \sum_{k=1}^{K}c_k \, \exp \bigl\{-\frac{1}{2}
	(x_i -\mu_k)^2/\sigma^2   \bigr\}
	\), $i=1,\ldots,n$, $j=1,\ldots,K$.
\end{enumerate}
\subsection{Asymptotic Distribution of $\bm{\beta}$}
\label{sec:asym}
First, assume that we place a normal prior on $\pmb{\beta}$ with mean $\pmb{\beta}_0$ and covariance matrix $A_0$. We assume that $A_0$ is known and is positive definite. Under certain regularity conditions, and as $n\rightarrow \infty$, the posterior distribution of $\bm{\beta}$ converges to normal, with mean $m_n$ and covariance $J_n$, where
\[
\begin{split}
J_n & = J(\hat{\bm{\beta}}_n) + A_0^{-1},\\
m_n & = J_n^{-1}\left[A_0^{-1}\bm{\beta}_0 +J(\hat{\bm{\beta}}_n) \bm{\hat{\beta}}_n \right]
\end{split}
\]
and $\hat{\bm{\beta}}_n$ is the maximum likelihood estimate of $\bm{\beta}$, $\bm{\beta}_0$ is the prior mean, and $J(\hat{\bm{\beta}}_n)$ is the negative second derivative of the log likelihood evaluated at $\hat{\bm{\beta}}_n$. A proof is presented in the Appendix. By placing a normal prior on $\bm{\beta$} with mean $\bm{0}$ and covariance matrix $\dfrac{1}{\tau^2}\bm{\beta}^{'}P^{*}\bm{\beta}$, the posterior distribution of $\bm{\beta}$ converges to normal, with mean $m_{n}$ and covariance $J_{n}$, where
\[
\begin{split}
J_{n}&=J(\hat{\bm{\beta}}_n) + \left(\dfrac{\bm{\beta}^{'}P^{*}\bm{\beta}}{\tau^2}\right)^{-1} \\
m_{n}&= J^{-1}_{n}J(\hat{\bm{\beta}}_n)\hat{\bm{\beta}}_n\\
J(\hat{\bm{\beta}}_n) &= \left( \dfrac{\partial^2}{\partial \bm\beta\partial \bm\beta^{'}}\sum_{j=1}^{K}n_j\log \left[    \dfrac{ \exp(\beta_j)}{\sum_{l=1}^{K}\exp(\beta_l) }\right]^{n_j} \right)_{\bm{\beta}=\bm{\hat{\beta}}_n}.
\end{split}
\]

\section{Dirichlet Process Mixture Models}
\label{sec:dpmm}
Unlike the mixture of Section~\ref{sec:mix_gauss}, in this section, mixtures with a countably infinite number of components are used by placing a Dirichlet process prior on the mixing proportions \citep{neal:2000}.
Dirichlet process mixture models (DPMM) were originally proposed by \cite{Ferguson1973}, \cite{antoniak:1974}
and \cite{ferguson:1983} and were first used in practice by \cite{escobar1994}.
Given data $x_1,\ldots,x_n$ independently drawn from some unknown distribution, a direct formulation of the DPMM
is as follows \citep{neal:2000}.
\renewcommand{\arraystretch}{1.5}
\begin{equation}
\begin{array}{rrl}
x_i \mid \theta_i & \sim & F(\theta_i) \\
\theta_i \mid G & \sim & G \\
G &\sim & DP(\alpha, G_0). 
\end{array}
\label{eq:dpmm}
\end{equation}
\renewcommand{\arraystretch}{1}
In (\ref{eq:dpmm}), $\theta_i$ (possibly a vector) is the parameter of the mixture component to which $x_i$ belongs, and $F$ is the distribution of the mixture components. For example, if $F$ is $N(\mu_i,\sigma_i^2)$, then
$\bm\theta_i=(\mu_i,\sigma_i^2)$. Two data points $x_i$ and $x_j$, belonging to the same component,
share the same component parameters, i.e., $\theta_i=\theta_j$.
The distribution $G$ is an infinite discrete distribution drawn from the Dirichlet process $DP(\alpha,G_0)$
with concentration parameter $\alpha$ and base distribution $G_0$. For the $i$th data point,
an atom $\theta_i$ is drawn from $G$, which may be equal to $\theta_j$ corresponding to $x_j$,
due to the discreteness of $G$. The Dirichlet process can be represented by the Stick Breaking scheme \citep{Sethuraman1994}, the Chinese Restaurant Process \citep{Aldous1985}, and the P\'olya Urn Scheme \citep{Blackwell1973}. In this work, we use the Stick Breaking representation. The stick breaking prior directly generates $\{\pi_k\}_{k=1}^\infty$ according to
\begin{equation}
%\begin{split}
%v_k & \sim  \mbox{beta}(1,\alpha) \\
\pi_1=v_1, \enskip \pi_k  =  v_k\prod_{\ell=1}^{k-1} (1-v_\ell), \enskip k\geq2, \\
%\end{split}
\label{eq:stick}
\end{equation}
where the random variables $v_1, v_2,\ldots$ are iid $\mbox{beta}(1,\alpha)$.
Having generated the $\pi_k$, $G$ can now be expressed as $G=\sum_{k=1}^\infty \pi_k \delta(\phi_k)$, where
$\phi_k \sim G_0$ are the distinct component parameters.
%%%%%%%%%%%%%%%%%%%%%%%%%%%%%%%%%%%%%%%%%%%%
\subsection{Infinite Mixture Models} 
Consider the finite mixture model
\renewcommand{\arraystretch}{1.5}
\[
\begin{array}{rrl}
x_i \mid z_i, \bm{\phi} & \sim & F(\phi_{z_i}) \\
z_i \mid \bm{\pi} & \sim & \mbox{Mult}(\pi_1,\ldots, \pi_K) \\
\phi_{z_i} & \sim & G_0 \\
\bm{\pi} & \sim & \mbox{Diriclet}(\alpha/K,\ldots,\alpha/K),
\end{array}
\]
\renewcommand{\arraystretch}{1}
where the $z_i$ are component indicators, $\bm{\phi}=(\phi_1,\ldots,\phi_K)'$, and $\bm{\pi}=(\pi_1,\ldots,\pi_K)'$.
\cite{neal:2000} shows that the limit of $p(z_{n+1} \mid z_1,\ldots, z_n)$, as $K\rightarrow\infty$, implies the
conditional probabilities for the $\theta_i$, which in turn shows the equivalence of the
infinite mixture model and the DPMM.
%%%%%%%%%%%%%%%%%%%%%%%%%%%%%%%%%%%%%%%%%%%%%%%
\subsection{Implementation}
In this paper, we use the DPMM with Gaussian components, i.e., model~(\ref{eq:dpmm}), where
$F(\theta_i)$ is $N(\mu_i,\sigma_i^2)$. More specifically, using the stick-breaking prior,
$\mbox{Stick}(\alpha)$, given in (\ref{eq:stick}), the model
can be written as
\renewcommand{\arraystretch}{1.5}
\[
\begin{array}{rrl}
x_i \mid z_i, \bm\phi & \sim & N(\mu_{z_i},\sigma_{z_i}^2),\enskip i=1,\ldots,n \\
z_i \mid \bm\pi & \sim & \mbox{Mult}(\bm\pi) \\
\bm\pi & \sim & \mbox{Stick}(\alpha) \\
\phi_{z_i} & \sim & G_0. \\
\end{array}
\]
\renewcommand{\arraystretch}{1}
We implement the MCMC scheme of \cite{Ishwaran2002}, who truncate $G=\sum_{k=1}^\infty \pi_k \delta(\phi_k)$
by $G_N=\sum_{k=1}^N \pi_k \delta(\phi_k)$, where $N$ is pre-specified. The priors on $(\mu_k,\sigma_k^2)$ and on $\alpha$ are as follows.
\begin{enumerate}
	\item
	$(\mu_k \mid \theta, \sigma_\mu^2) \stackrel{iid}{\sim} N(\theta,\sigma_\mu^2)$, $\sigma_\mu^2$ is fixed.
	\item
	$(\sigma_k^2 \mid \nu_1,\nu_2) \stackrel{iid}{\sim} IG(\nu_1,\nu_2)$, $\nu_1$, $\nu_2$ are fixed.
	\item
	$(\alpha \mid \eta_1,\eta_2) \sim G(\eta_1,\eta_2)$, $\eta_1$, $\eta_2$ are fixed.
	\item
	$\theta \sim N(0,A)$, $A$ is fixed.
\end{enumerate}
For the sampling scheme, see \cite{Ishwaran2002}.

\section{Simulation and Example}
\label{sec:example}
\subsection{Simulations}
This section presents a comparison of the three density estimation methods described in sections 2-4, along with the kernel and log-spline densities. For the kernel density, we use two different bandwidths, which are UCV and SJ. UCV implements unbiased cross-validation, whereas SJ implements the methods of \cite{Sheather1991} to select the bandwidth using pilot estimation of derivatives. We simulate data from five different distributions taken from \cite{schellhase:kauermann:2012} (Table \ref{tab:0}). As in \cite{schellhase:kauermann:2012}, we use two different sample sizes: $n=100$ and $n=400$. For our PGM approach, we use three different values of $K$: 20, 30, and 50. For the DPPM method, we set the number of mixture components equal to 35.

\begin{table}
\begin{center} 
	\begin{tabular}{ccc}
		\hline
		pdf && Distribution  \\ \hline
	$f_1(x)$	&& $N\left(0,1\right)$  \\
		$f_2(x)$ && $\dfrac{1}{2}N\left(-\dfrac{1}{2},\dfrac{1}{4}\right) +\dfrac{1}{2}N\left(\dfrac{1}{2},\dfrac{1}{4}\right)$ \\
		$f_3(x)$&& $\dfrac{1}{2}N\left(-\dfrac{3}{2},1\right) + \dfrac{1}{2}N\left(\dfrac{3}{2},1\right)$ \\
		$f_4(x)$&& $\dfrac{13}{20}N\left(-1,\dfrac{1}{2}\right)+\dfrac{2}{20}N\left(-\dfrac{1}{2},\dfrac{1}{2}\right)+\dfrac{1}{20}N\left(0,1\right)+\dfrac{3}{20}N\left(\dfrac{1}{2},\dfrac{1}{2}\right)+\dfrac{1}{20}N\left(1,\dfrac{1}{2}\right)$		     \\
		$f_5(x)$&& $\text{Gamma}\left(3,1\right)$  \\ \hline
	\end{tabular}
\end{center}
	\caption{Distributions from \cite{schellhase:kauermann:2012}.}\label{tab:0}
\end{table}

To sample $\bm{\beta}, \tau^2, a, \bm{z}$ in PGM, we use a block of HMC and Gibbs sampler. The HMC technique requires fewer iterations to explore the parameter space and converges rapidly to the target distribution \citep{Hartmann2017}. Therefore, we implement the HMC and Gibbs sampler with 5000 iterations and 1000 burn-in. HMC has two tuning parameters: the step size $v$ and the number of leapfrog steps $L$. We use trial and error to set their values. In particular, we select $v=0.018$ and $L=10$. For Lindsey and DPPM methods, we implement the Gibbs sampler. We measure the performance of the estimates using the integrated mean squared error (IMSE). The IMSE is obtained by
\[
IMSE = \dfrac{1}{s}\sum\limits_{l=1}^{s}\Bigl{\{ }  MSE(\hat{f})   \Bigr{\} },
\]
where $s$ is the number of samples drawn from each distribution, which in our simulation study is set to 100. The MSE measures the difference between the true density $f$ and its estimate and is given by

\[
MSE = \dfrac{1}{n} \sum_{j=1}^{n} \left \{ f(x_j) - \hat{f}(x_j) \right\}^2,
\]
where $f(x_j)$ and $\hat{f}(x_j)$  are the true density and its estimate evaluated at $x_j$, and $n$ is the sample size. The \textit{logspline} and the \textit{kernel density} estimates are estimated using the logspline package \citep{logspline} and stats package \citep{stats}, respectively.

\begin{figure}[H]
	\begin{minipage}{0.5\textwidth}
		\centering
		\includegraphics[width=1.0\linewidth]{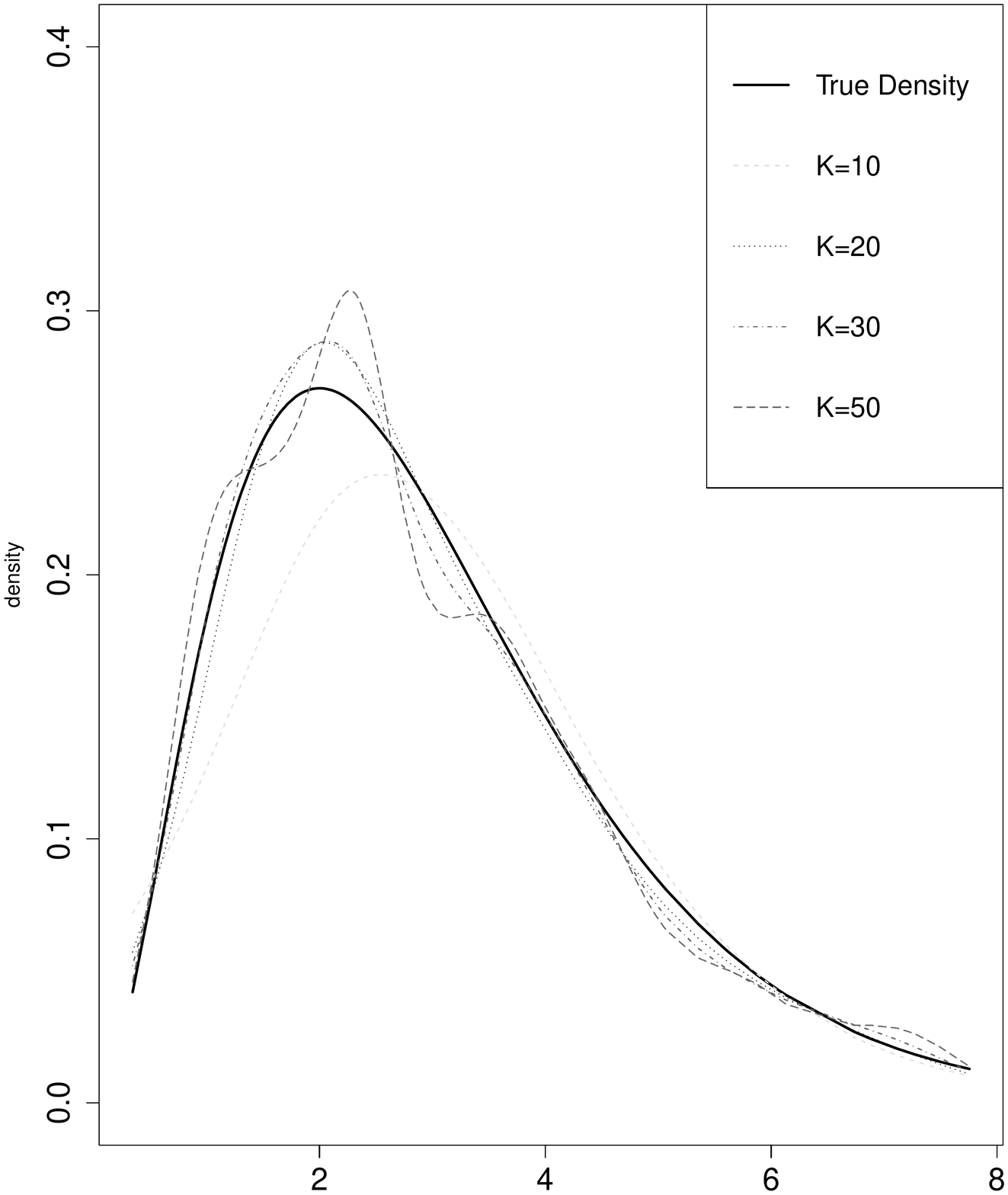}
	\end{minipage}\hfill
	\begin {minipage}{0.5\textwidth}
	\centering
	\includegraphics[width=1.05\linewidth]{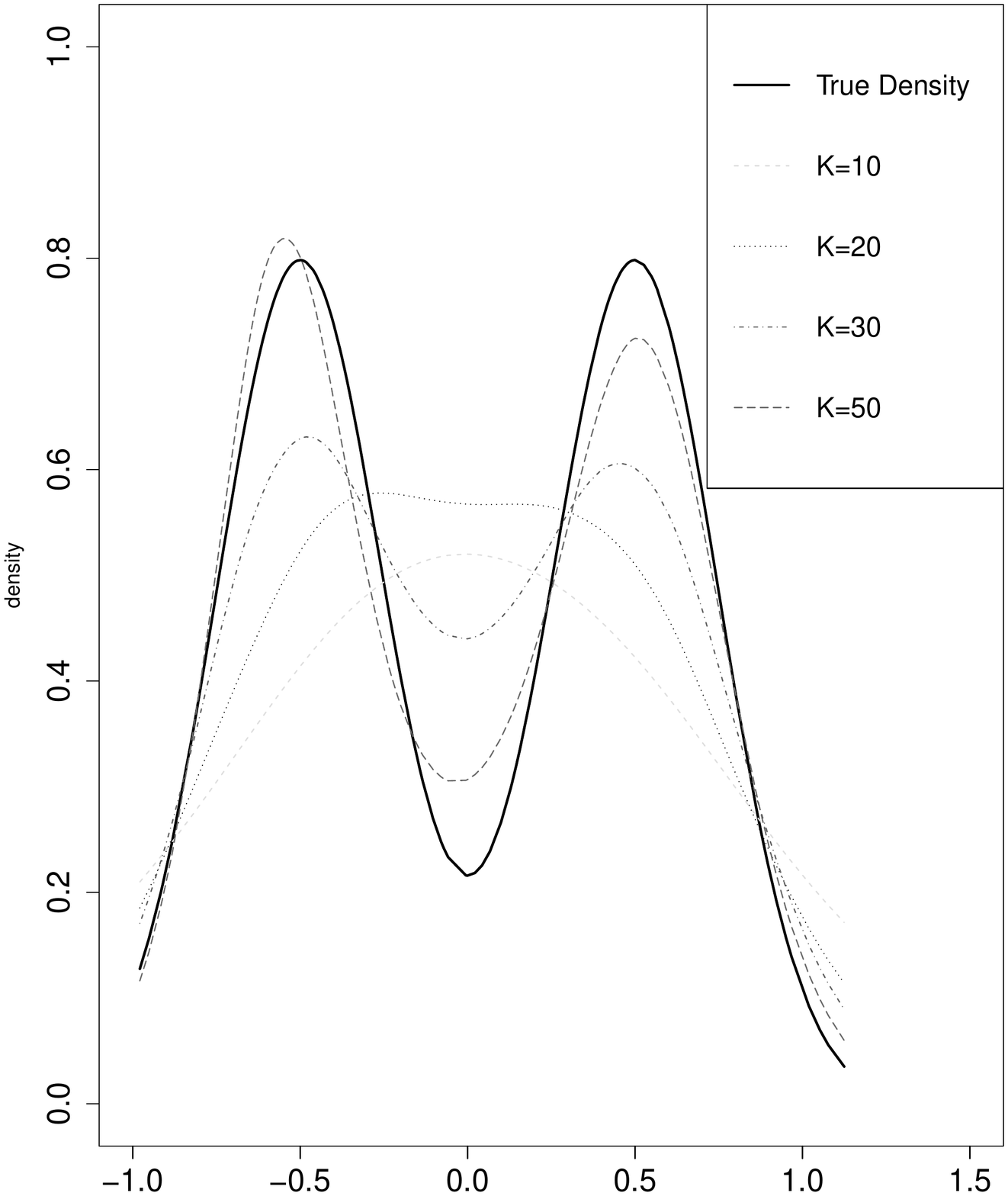}
	\caption{The true densities $f_5(x)$ (left graph) and $f_2(x)$ (right graph) along with PGM estimates using different values of $K$. In both cases, the sample size is $n=200$.}\label{Fig:density}
\end{minipage}
\end{figure}

Table~\ref{tab:1} presents the results of the simulations for the PGM, LM, DPPM, Kernel densities, and log-spline methods. Our Bayesian implementation of PGM and Lindsey methods appear to perform promisingly well in comparison with the DPMM, Kernel, log-spline methods, and the results from \cite{schellhase:kauermann:2012} and in some cases even better than its frequentist counterpart. Figure \ref{Fig:density} displays the PGM estimates, based on a single sample from $f_5(x)$ and $f_2(x)$, respectively. $f_5(x)$ is the Gamma distribution with shape 3 and scale 1, and $f_3(x)$ is the mixture of two components. Increasing the value of $K$ improves the fit.

\begin{table}
\begin{center}
\begin{tabular}{ c c c c c c c c c c}
\hline
pdf & Sample size &  & PGM & &LM&DPMM &\multicolumn{2}{c}{Kernel} & Log-spline\\
\cline{3-5} \cline{8-9}
&             &  K=20	  & K=30 	& K=50  &	&	 & bw=UCV	&bw=SJ&	  \\
\hline
$f_1(x)$  &$n=100$  &1.106 &1.841 &2.952 &1.393 &0.791 &1.486&1.015&6.462 \\
          &$n=400$  &0.306 &0.535 &0.910 &0.275 & 0.157&0.369&0.303&1.224 \\   
$f_2(x)$  &$n=100$  &50.974&25.824&6.838 &45.711&7.326 &7.702&6.015&25.357\\
          &$n=400$  &42.178&15.642&2.365 &38.100&1.591 &2.990&2.468&6.687 \\
$f_3(x)$  &$n=100$  &0.532 &0.534 &0.913 &0.805 &1.278 &0.475&0.557&1.777 \\       
		  &$n=400$  &0.220 &0.196 &0.320 &0.286 &0.288 &0.203&0.226&0.485 \\
$f_4(x)$  &$n=100$  &10.061&24.875&31.106&22.358&35.321&2.808&2.240&9.962 \\
		  &$n=400$  &1.474 &2.322 &5.440 &19.815&30.593&0.829&0.699&2.344 \\
$f_5(x)$  &$n=100$  &1.004 &1.081 &1.440 &1.676 & 1.634&1.658&1.484&2.870 \\
		  & $n=400$ &0.536 &0.336 &0.407 &0.205 & 0.627&0.209&0.58 &0.706 \\
\hline
\end{tabular}
\end{center}
\caption{Comparison of the IMSE values $(\times10^3)$ based on 100 simulated samples.}\label{tab:1}
\end{table}

\subsection{Example: Daily Returns.}
\begin{figure}[H]
\centering
\includegraphics[width=1\linewidth]{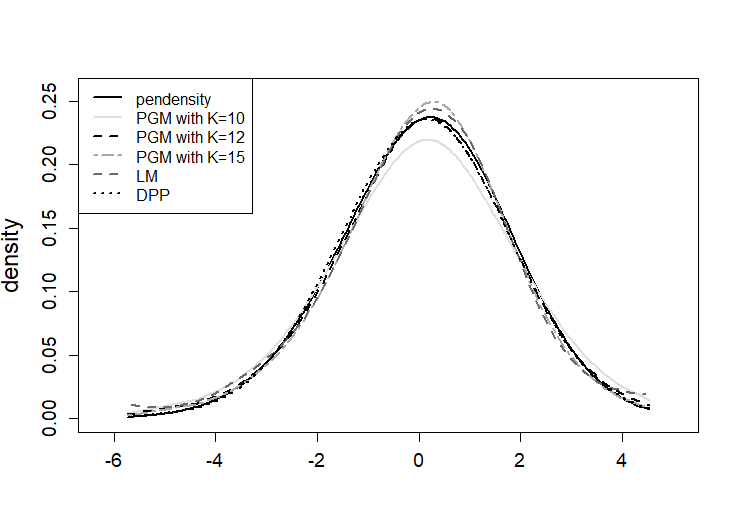}
\caption{Density estimates of the return of the Allianz AG in 2006 by penalized mixture approach (pendensity), PGM method using three different values of $K$, LM, and DPP methods.}\label{Fig:density2}
\end{figure}

We use the daily return data set from \cite{schellhase:kauermann:2012} presented in Section 3.2 of their paper, which represents the return of the two German stocks Deutsche Bank AG and Allianz AG in 2006. The corresponding density estimates of the Bayesian penalized mixture approach is given in Figure \ref{Fig:density2}. We show the PGM estimate along with 3 competitors, which are Lindsey's method, Dirichlet process mixture model, and penalized mixture approach (pendensity) \citep{schellhase:kauermann:2012}. For the Bayesian PGM method, we used three different values for $K$: 10, 12, and 15. We implement the MCMC scheme with 5000 iterations, the first 1000 of which are used as burn-in. The corresponding estimates are displayed in Figure \ref{Fig:density2}. The PGM with $K=12$ and the DPPM approaches provide almost identical density estimates compared to the pendensity method.

\section{Summary}
\label{sec:summary}
In this paper, we have derived a nonparametric Bayesian approach for Lindsey \citep{Lindsey1974} and penalized Gaussian mixtures method \citep{Ghidey2004}. Lindsey's method recasts density estimation as a regression problem where we implement the cubic smoothing spline to estimate the regression function. For the PGM method, we used the HMC algorithm to estimate the weights mixtures. Moreover, we showed, under certain regularity, that the posterior distribution of the estimates in PGM converges to a Normal distribution. We ran simulations and compared our approaches to DPPM, Kernel, and log-splines methods. In general, it appeared from simulations that our approaches performed promisingly well when compared to the competitors, and in some cases even better than its frequentist counterpart. Among the Bayesian methods, the Lindsey approach is computationally less expensive.
\vfill
\clearpage

\section*{Acknowledgement}
The authors were supported in part by  the National Security Agency
under Grant Number H98230-12-1-0246. The United States Government is authorized to reproduce and distribute reprints not-withstanding any copyright notation herein.

\section*{Appendices}

\noindent\textbf{Appendix A1. Sampling from $p(\bm\beta \mid \tau^2,\bm z)$ of Section~\ref{sec:mix_gauss}}\medskip

To draw from the conditional posterior density (\ref{eq:cond_beta}), we use the HMC algorithm \citep{Neal2011}. We first give an overview of HMC and then apply it to (\ref{eq:cond_beta}). HMC, also known as Hybrid Monte Carlo, is an MCMC method to generate posterior samples for which direct sampling is difficult. It uses the gradient of the posterior density and the Hamiltonian system to sample successive states for the Metropolis-Hastings algorithm with a large acceptance probability. The underlying logic of the HMC sampling is as follows. To sample from a posterior distribution $\pi(\bm\beta|\bm x)$, we treat the parameter $\bm\beta$ as a particle and denote its value at its current position. We define the potential energy and the kinetic energy, respectively, as
\[
\begin{split}
U(\bm\beta) &= - \log \{\pi(\bm\beta|\bm x) \}\\
K(\bm u) &= \dfrac{1}{2}\bm u^TM^{-1}\bm u,
\end{split}
\]
where $\bm u=(u_1,\;u_2,\cdots,u_s)^T$ is the momentum vector, and  $M$ is the mass matrix, also known as the dispersion matrix. The kinetic energy $K(\bm u)$ arises from the Gaussian distribution $N(0,\;M)$, where $M$ is a symmetric positive definite matrix. We set $M$ equal to the identity matrix. The Hamiltonian system is defined as 
\[ H(\bm\beta,\;\bm u) = U(\bm\beta)+ K(\bm u). \] 
The position of $\bm\beta$ and the momentum $\bm u$ of the particle change over time and are determined by the partial derivatives of the Hamiltonian system. These partial derivatives give rise to the so-called Hamiltonian equations of motion
\begin{equation}
\begin{split}
\dfrac{d \bm\beta}{d t}=&\dfrac{\partial H}{\partial \bm u} =M^{-1}\bm u,\\
\dfrac{d \bm u}{d t}=- \dfrac{\partial H}{\partial \bm\beta}=&-\dfrac{\partial U(\bm\beta)}{\partial \bm\beta} =\dfrac{\nabla \pi(\bm\beta|x)}{\pi(\bm\beta|x)}.
\end{split}
\label{eqHMC1}
\end{equation}
\cite{Neal2011} showed that these Hamiltonian equations are reversible, invariant, and volume-preserving, which makes the Hamiltonian system suitable for MCMC sampling schemes. When $\pi{(\bm\beta|x)}$ lacks a closed form, equations (\ref{eqHMC1}) have no analytic solutions. Thus, the solution is approximated at discrete time steps. Following \cite{Neal2011}, we apply the leapfrog integration method to approximate the solution of the Hamiltonian equations. First, a small step size $\vartheta$ is selected, and the starting value of $\bm \beta$ is the maximizer of $p(\bm\beta \mid \tau^2,\bm z)$. Then, given the current value of $\bm\beta$ and $\bm u$ at time $t$, the position and momentum at time $t+\vartheta$ are updated as follows.
\[
\begin{split}
\bm u\left( t+\dfrac{1}{2}\vartheta \right) &=\bm u(t)-\dfrac{1}{2}\vartheta\dfrac{\partial U(\bm\beta(t))}{\partial \bm\beta}\\
\bm\beta \left(t+\vartheta \right)&=\bm\beta(t)+ \vartheta M^{-1}\bm u \left(t+\dfrac{1}{2}\vartheta \right)\\
\bm u\left(t+\vartheta \right) &= \bm u \left( t+ \dfrac{1}{2}\vartheta\right) -\dfrac{1}{2}\vartheta \dfrac{\partial U(\theta(t+\vartheta)) }{ \partial \theta}.
\end{split}
\]
Sometimes the approximation introduces errors, and an accept-reject algorithm is required to conserve the invariant property of HMC \citep{Neal2011}. The procedure works as follows. In the first step, new values for the momentum vector $\bm u$ are randomly drawn from a Gaussian distribution $N(0,\;M)$, independently of the current values of $\bm\beta$. In the second step, starting with the current state, $(\bm\beta, \bm u)$, a Hamiltonian system is simulated for $L$ steps using the leapfrog method, with a step size of $\vartheta$. At the end of this $L$-step trajectory, the proposed state $(\bm\beta^*,\bm u^*)$ is accepted with probability 
\[
\min\left[ 1,\exp \left(-H(\bm\beta^*,\bm u^*) + H(\bm\beta, \bm u) \right) \right] = \min\left[ 1,\exp \left( -U(\bm\beta^*) + U(\bm\beta) -K(\bm u^*) +K(\bm u)\right) \right],
\label{eqHmc2} 
\]
where $U(\bm\beta) = -\log(\pi(\bm\beta|x))$ and $K(\bm u)=\dfrac{1}{2}\bm u^TM^{-1}\bm u$. If the proposed state is rejected, the next state is the same as the current one. To apply HMC to the sampling of $\bm{\beta}$ in our case, we need to obtain $-\log p(\bm{\beta} \mid \tau^2,\bm{z})$ and its gradient.
\[
\begin{split}
-\log p(\bm{\beta} \mid \tau^2,\bm{z}) &= -\sum_{j=1}^{K} n_j\log c_j +\dfrac{1}{2\tau^2}\bm{\beta}^{'}P^{*}\bm{\beta}\\
-\dfrac{\partial \log p(\bm{\beta} \mid \tau^2,\bm{z})}{\partial \bm{\beta}} &=-\begin{pmatrix}
n_2 -n c_2 \\ 
n_3 -n c_3\\  
\vdots \\
n_{K} -n c_{K}\\ 
\end{pmatrix}  +\frac{1}{\tau^2}P^*\bm\beta.
\end{split}
\]

\noindent\textbf{Appendix A2. Proof of the asymptotic Distribution of Section~\ref{sec:asym}} \\

The posterior distribution of $\bm{\beta}$ is\medskip

	\[
	\begin{split}
	\pi\left( \bm{\beta}|X,\bm{y}\right)&\propto \pi(X,\bm{y}|\bm{\beta})\pi\left(\bm{\beta} \right).\\
	& \propto \exp\left( \log\prod_{i=1}^{n}   c_{z_i} -\dfrac{1}{2}(\bm{\beta}-\bm{\beta_0} )^T A_0^{-1}(\bm{\beta}-\bm{\beta_0} )     \right)\\
	& = \exp\left( \log\prod_{j=1}^{K}   c_{j}^{n_j} -\dfrac{1}{2}(\bm{\beta}-\bm{\beta_0} )^T A_0^{-1}(\bm{\beta}-\bm{\beta_0} )     \right)\\
	&=\exp\left( \log\pi(X,\bm{y}|\bm{\beta}) -\dfrac{1}{2}(\bm{\beta}-\bm{\beta_0} )^T A_0^{-1}(\bm{\beta}-\bm{\beta_0} ) \right),\\
	\end{split}
	\]
	where
	\[
	\pi(X,\bm{y}|\bm{\beta}) = \prod_{j=1}^{K}   c_{j}^{n_j}\text{ and } c_j=\dfrac{\exp(\beta_j)}{\sum_{l=1}^{K}\exp{\beta_l}}.
	\]
	Similar to \cite{Bernardo1994}, we expand the logarithm term about its maximum $\bm{\hat{\beta}}_n$, obtained by setting the first derivatives of the logarithm to zero
	\[
	\log\pi(\bm{\beta}|X,\bm{y}) = \log\pi(X,\bm{y}|\bm{\hat{\beta}}_n)-\dfrac{1}{2}(\bm{\beta}-\bm{\hat{\beta}}_n )^T A_0^{-1}(\bm{\beta}-\bm{\hat{\beta}}_n ) +R_n,
	\]
	where $R_n$ is the remainder, which is small for large $n$.
	\newline
	In addition, we have:  $\log\left(\pi(X,\bm{y}|\bm{\hat{\beta}}_n)\right)=   
	-\dfrac{1}{2}\left(\bm{\beta} -\bm{\hat{\beta}}_n\right)^TJ(\bm{\hat{\beta}}_n)\left(\bm{\beta} -\bm{\hat{\beta}}_n\right)$, where 
	\[
	J(\bm{\hat{\beta}}_n)= \left(-\dfrac{\partial^2 \log\pi(X,\bm{y}|\bm{\beta})}{\partial\bm\beta \partial\bm\beta^{'}} \right)_{\bm{\beta}=\bm{\hat{\beta}}_n}.
	\]
	If we assume $n$ is large and ignore constants of proportionality, we have:
	\[
	\begin{split}
	\pi(\bm{\beta}|X,\bm{y}) &\propto \exp \left\lbrace  
	-\dfrac{1}{2}\left(\bm{\beta} -\bm{\hat{\beta}}_n\right)^TJ(\bm{\hat{\beta}}_n)\left(\bm{\beta} -\bm{\hat{\beta}}_n\right) -\dfrac{1}{2}(\bm{\beta}-\bm{\beta_0} )^T A_0^{-1}(\bm{\beta}-\bm{\beta_0} ) 
	\right\rbrace\\
	&=\exp \left\lbrace  
	-\dfrac{1}{2}\left( \left(\bm{\beta} -\bm{\hat{\beta}}_n\right)^TJ(\bm{\hat{\beta}}_n)\left(\bm{\beta} -\bm{\hat{\beta}}_n\right) +(\bm{\beta}-\bm{\beta_0} )^T A_0^{-1}(\bm{\beta}-\bm{\beta_0} ) 
	\right)\right\rbrace\\
	&=\exp \left\lbrace
	-\dfrac{1}{2} \left(
	\bm{\beta^T}J(\bm{\hat{\beta}}_n)\bm{\beta} -  \bm{\beta^T}J(\bm{\hat{\beta}}_n)\bm{\hat{\beta}}_n -\bm{\hat{\beta}}_nJ(\bm{\hat{\beta}}_n)\bm{\beta} +
	\bm{\hat{\beta}}_nJ(\bm{\hat{\beta}}_n)\bm{\hat{\beta}}_n
	+\bm{\beta^T}A_0^{-1}\bm{\beta}-\bm{\beta^T}A_0^{-1}\bm{\beta_0}
	\right)
	\right\rbrace\\
	& \;\;+\exp \left\lbrace  
	-\dfrac{1}{2}\left( 
	-\bm{\beta_0^T}A_0^{-1}\bm{\beta} + \bm{\beta_0^T}A_0^{-1}\bm{\beta_0}
	\right)
	\right\rbrace\\
	&\propto  \exp \left\lbrace  
	-\dfrac{1}{2}\left(
	\bm{\beta^T}\left[J(\bm{\hat{\beta}}_n)+A_0^{-1}\right]\bm{\beta}-
	2\bm{\beta^T}J(\bm{\hat{\beta}}_n)\bm{\hat{\beta}}_n +2\bm{\beta^T}A_0^{-1}\bm{\beta_0}
	\right)
	\right\rbrace\\
	&  =  \exp \left\lbrace  
	-\dfrac{1}{2}\left(
	\bm{\beta^T}\left[J(\bm{\hat{\beta}}_n)+A_0^{-1}\right]\bm{\beta}-2\bm{\beta^T}\left[J(\bm{\hat{\beta}}_n)\bm{\hat{\beta}}_n+A_0^{-1}\bm{\beta_0} \right]
	\right)
	\right\rbrace.\\
	\end{split}
	\]
	Setting $ J_n  = J(\hat{\bm{\beta}}_n) + A_0^{-1}$  and 
	$m_n  = J_n^{-1}\left[A_0^{-1}\bm{\beta_0} +J(\hat{\bm{\beta}}_n) \bm{\hat{\beta}}_n \right]$, we have:
	\[
	\pi(\bm{\beta}|X,\bm{y}) \propto\exp \left\lbrace -\dfrac{1}{2}\left(
	\bm{\beta^T}J_n\bm{\beta} -2\bm{\beta^T}J_nm_n \right)
	\right\rbrace.
	\]
	We complete the square above by adding and subtracting  $m_n^TJ_nm_n$. Therefore,
	\[
	\pi(\bm{\beta}|X,\bm{y}) \propto
	\exp \left\lbrace  
	-\dfrac{1}{2}\left(
	\left[\bm{\beta} -m_n  \right]^TJ_n\left[\bm{\beta} -m_n  \right]           
	\right)
	\right\rbrace,
	\]
	is the kernel of $N_p\left( m_n,\;J_n\right)$, with $m_n$ and $J_n$ defined above.
	\newpage

\noindent\textbf{Appendix A3. The effect of the value A on the weights}
 \\
 
\begin{figure}[H]
	\centering
	\includegraphics[width= 1\linewidth, height=0.68\textheight]{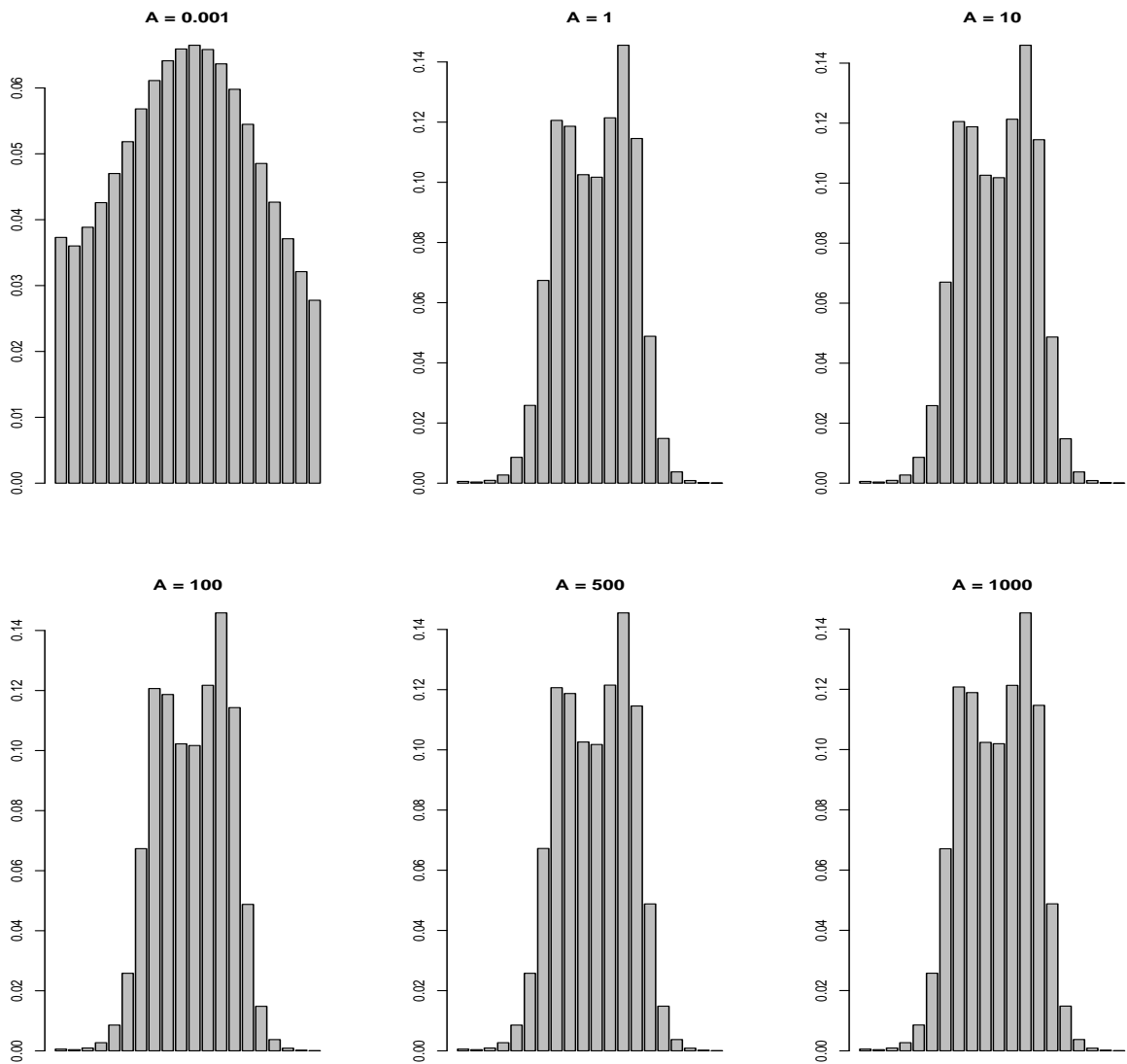}
	\caption{The effect of the prior on the weights for different values of A.}\label{Fig:Avalues}
\end{figure}

\noindent In this example, we study the effect of the value A on the weights, $c_j = \exp(\beta_j)/\sum_{h=1}^{K} \exp(\beta_h)$, that use the logistic transformation such that $j=1,\cdots,K$ and $\sum_{j=1}^{K}c_j=1$. We simulate datasets with $n=400$ from the following model: $
\dfrac{1}{2} N\left(-\dfrac{3}{2},1\right) + \dfrac{1}{2}N\left(\dfrac{3}{2},1\right).$ We set $A=\left(1e{-03},\; 1,\; 10,\; 100,\; 500,\; 1000\right)$ and $k=20$. Similarly, we use a block of HMC and Gibbs sampler with 5000 iterations and 1000 burn-in. Figure \ref{Fig:Avalues} depicts the bar plots of the weights $c_j$'s based on different values of $A$. We can see the absence of the effect of the prior on the weights after the logistic transformation; except when $A$ is near to zero where the weights take values  near to $1/K$.

	\bibliographystyle{asa}
	\bibliography{references}
\end{document}